%% file: arXiv.tex
\documentclass{article}

\usepackage[english]{babel}
\usepackage[letterpaper,top=2cm,bottom=2cm,left=3cm,right=3cm,marginparwidth=1.75cm]{geometry}
\usepackage{amsmath}
\usepackage{graphicx}
\usepackage{verbatim}
\usepackage[colorlinks=true, allcolors=blue]{hyperref}
\usepackage{dsfont}
\usepackage{theorem}
\usepackage{float}
\usepackage{caption}
\usepackage[authoryear]{natbib}
\usepackage{listings}
\usepackage{enumerate}
\usepackage{rotating}
\usepackage{amssymb}

\def \dsR {\text{$\mathds{R}$}}

\theoremstyle{plain}
\newtheorem{theorem}{Theorem}[section]

\newtheorem{proposition}[theorem]{Proposition}
\newtheorem{corollary}[theorem]{Corollary}

\theoremstyle{remark}
\newtheorem{definition}[theorem]{Definition}
\newtheorem{example}[theorem]{Example}
\newtheorem{remark}[theorem]{Remark}

\title{Anisotropic multidimensional smoothing using Bayesian tensor product P-splines}
\author{Paul Bach and Nadja Klein\footnote{The authors gratefully acknowledge support by the German research foundation (DFG) through the  Emmy Noether grant KL 3037/1-1.}\\
Humboldt-Universität zu Berlin}

\begin{document}
\maketitle

\begin{abstract}
\input{abstract}

\end{abstract}

\vspace{1em}

\noindent%

\vspace{1em}

\input{main}

\bibliographystyle{dcu}
\bibliography{references}

\end{document}

%% file: abstract.tex
We introduce a highly efficient fully Bayesian approach for anisotropic multidimensional smoothing. The main challenge in this context is the Markov chain Monte Carlo update of the smoothing parameters as their full conditional posterior comprises a pseudo-determinant that appears to be intractable at first sight. As a consequence, most existing implementations are computationally feasible only for the estimation of two-dimensional tensor product smooths, which is, however, too restrictive for many applications. In this paper, we break this barrier and derive closed-form expressions for the log-pseudo-determinant and its first and second order partial derivatives. These expressions are valid for arbitrary dimension and very efficient to evaluate, which allows us to set up an efficient MCMC sampler with adaptive Metropolis-Hastings  updates for the smoothing parameters. We investigate different priors for the smoothing parameters and discuss the efficient derivation of lower-dimensional effects such as one-dimensional main effects and two-dimensional interactions. We show that the suggested approach outperforms previous suggestions in the literature in terms of accuracy, scalability and computational cost and demonstrate its applicability by consideration of an illustrating temperature data example from spatio-temporal statistics.

%% file: main.tex
\section{Introduction}
There are numerous settings in statistics where measurements $(x_i,y_i)\in\dsR^{p}\times \dsR,\ i=1,\dots,n, $ $p\geq 2$ are available and a smooth surface estimate $\widehat{f}(x)$ with varying degree of smoothness in each dimension $1\leq j\leq p$ is required. One  example which we also use for illustration later on is from spatio-temporal statistics: Here, the $y_i$ are noisy temperature measurements and the $x_i$ contain  spatio-temporal information about these measurements. A smooth surface estimate allows one to predict the temperature at locations and time points where no measurements are available and to gain general insights into the spatio-temporal temperature dynamics. For this example, it is highly desirable to allow not only for a different amount of smoothing for the temporal dimension but also across the two spatial dimensions. This is because the temperature profile cannot necessarily be assumed to be comparably smooth in the north-south direction (across different latitudes) as in the east-west direction (across different longitudes) due to varying climatological gradients.

A general key distinction in the context of multidimensional smoothing is that between isotropic and anisotropic smoothing: Isotropic smoothing means that there is a single smoothing parameter and that every coordinate receives the same amount of smoothing. Anisotropic smoothing, in contrast, means that there are $p=\text{dim}(x_i)$ smoothing parameters and that every coordinate receives its own amount of smoothing. The latter is generally desirable but much more challenging from a computational point of view. 

Until recently, the popular Bayesian P-splines approach of \cite{LanBre2004} has been limited to  isotropic smoothing. The main challenge to achieve  anisotropic smoothing is the Markov chain Monte Carlo (MCMC) update of the smoothing parameters. This is because their full conditional posterior comprises a pseudo-determinant that appears to be intractable at first sight. 

 Existing fully Bayesian approaches in the literature are unsatisfactory, either because of prohibitive runtimes or because they only allow for partially anisotropic smoothing: 
\cite{Woo2016} introduced the function \texttt{jagam}, which allows for a seamless combination of the \texttt{R} package \texttt{mgcv} \citep{wood2012mgcv} and the general purpose MCMC sampler \texttt{JAGS} \citep{PluHorLei2003b}. This approach works well for a two-dimensional tensor product smooth but it becomes extremely slow for dimension three or higher. The \texttt{R} package \texttt{bamlss} by \cite{UmlKleZei2018} also allows for anisotropic Bayesian smoothing and has e.g.~been applied by \cite{KohUmlGre2018} to estimate a two-dimensional tensor product smooth in a biomedical context. However, \texttt{bamlss} uses slice sampling with a stepping-out procedure \citep{Nea2003} to update the smoothing parameters. Similar to \texttt{jagam}, this becomes extremely slow for dimension three or higher. \cite{KneKleLan2019} introduced an alternative approach that relies on a discrete anisotropy parameter. This approach is implemented in \texttt{BayesX} \citep{BreKneLan2005} and much faster than those of \texttt{bamlss} or \texttt{jagam} for a three-dimensional smooth. However, the approach breaks down for a four-dimensional smooth and, in addition to that, it only allows for partially anisotropic smoothing. \cite{KneKleLan2019} partition the coordinates into two groups which are both treated isotropically. This leads to inferior performance in simulations but is also unsatisfactory from a practical perspective. In a spatio-temporal context, for instance, the approach allows to treat space and time anisotropically but it does not allow for a different amount of smoothing across all spatial dimensions. 

To the best of our knowledge, \texttt{Stan} \citep{CarGelHof2017} currently also does not offer a satisfactory solution: Approaches that implement tensor product P-splines using the \texttt{mgcv} constructor \texttt{te} do not seem to be readily available. The popular \texttt{R} package \texttt{rstanarm} \citep{GooGabAli2022}, for instance, only supports the alternative constructor \texttt{t2} based on \cite{WooSchFar2013} which uses a different roughness penalty. \cite{WooSchFar2013} have shown that the alternative penalty is comparable in terms of MSE. We can confirm this result but we found that \texttt{rstanarm} becomes unreliable for a three-dimensional tensor product smooth and extremely slow for dimension four or higher. 

The lack of efficient fully Bayesian approaches for anisotropic multidimensional smoothing stands in sharp contrast to tensor product spline smoothers that use restricted maximum likelihood (REML) for the selection of the smoothing parameters. Several efficient approaches have been developed \citep{Woo2011,RodLeeKne2015,WooFas2017} and are readily available in \texttt{R} packages such as \texttt{mgcv}. The fully Bayesian approach, however, has the advantage that the uncertainty of the variance parameters is taken into account in the estimation process. In addition to that, it is relatively straightforward to incorporate various complications such as heteroscedasticity or missing data into the fully Bayesian approach \citep[cf.][Section 6.9]{HarRupWan2018}.

In this paper, we close this gap and introduce a highly efficient fully Bayesian approach for anisotropic multidimensional smoothing. To overcome the obstacle posed by the pseudo-determinant we exploit a special representation of the overall roughness penalty matrix. This representation is closely related to the mixed model representation of tensor product smooths \citep{Woo2006b,LeeDur2011,RodLeeKne2015} and allows us to derive closed-form expressions for the log-pseudo-determinant and its partial derivatives. These expressions are very fast to evaluate which allows us to set up an efficient MCMC sampler with adaptive Metropolis Hastings (MH) proposals for the smoothing parameters. In summary, our work makes the following major contributions.
\begin{itemize}
    \item We introduce a highly efficient fully Bayesian approach for anisotropic multidimensional smoothing using Bayesian tensor product P-splines. Our approach allows for a different amount of smoothing for every coordinate and works well in estimating a function that depends on up to five continuous coordinates. 
    \item We derive efficient and adaptive MH updates for the smoothing parameters and show that our resulting algorithm outperforms previous suggestions in the literature by means of simulations: It is much faster and yields better performance in terms of mean squared error (MSE).
    \item Our approach can e.g.~be applied in a spatio-temporal context, where it allows for a different amount of smoothing for time and across all spatial dimensions. We demonstrate its applicability by consideration of a temperature data set with $n=12,672$ observations in a three-dimensional space-time setting.
\end{itemize}
The remainder of this paper is organized as follows: In Section~\ref{sec:AnisotropicPSplineModel} we introduce anisotropic multidimensional smoothing using Bayesian P-splines, whereas Section~\ref{sec:EfficientMCMC} details  our new approach for efficient posterior sampling. In Section~\ref{sec:MainEffectsAndInteractions} we discuss the efficient derivation of lower-dimensional effects such as one-dimensional main effects and two-dimensional interactions. Section~\ref{sec:EmpiricalEvidence} presents empirical evidence and Section~\ref{sec:Discussion} concludes with a discussion. The Appendix contains further technical details, proofs of our theoretical results and further background information for the temperature data set.

\section{Bayesian Anisotropic P-Spline Model}\label{sec:AnisotropicPSplineModel}
Throughout, we consider the $p$-dimensional nonparametric regression model
\begin{align}\label{NonparametricRegressionModel}
    y_i=f(x_i)+\epsilon_i,\ \epsilon_i\sim N_1(0,\sigma^2),\ i=1,\dots,n,
\end{align}
where $f:[0,1]^p\longrightarrow \dsR$ is an unknown function to be estimated and the $x_i$ are assumed to lie in the $p$-dimensional unit cube $[0,1]^p$ (without loss of generality). We assume that $f$ can be approximated by tensor product splines, i.e. 
\begin{align}\label{BSplineExpansion}
    f\approx \sum_{j=1}^D B_jb_j
\end{align}
for some unknown coefficient vector $b\in\dsR^D$, where the $B_j,\ j=1,\dots,D,$ are tensor product B-splines of the form
\begin{align*}
    B_1(x)=\prod_{j=1}^p B_1(x_j),\ B_2(x)=\prod_{j=1}^{p-1} B_1(x_j) B_2(x_p),\ \dots\ ,\ B_D(x)=\prod_{j=1}^p B_{d_j}(x_j),\quad x\in[0,1]^p.
\end{align*}
Thereby, the $p$ marginal bases $\{B_1(x_1),\dots,B_{d_1}(x_1)\},\dots,\{B_1(x_p),\dots,B_{d_p}(x_p)\}$ are cubic B-spline bases of dimensions $d_j,\ j=1,\dots,p,$ each covering the unit interval $[0,1]$. Following the Bayesian P-splines approach of \cite{LanBre2004}, we use a relatively large number of equidistant spline knots for the marginal B-spline bases. 
To prevent overfitting, we endow the tensor product B-spline coefficient vector $b\in\dsR^D$ with a smoothness prior that encourages a smooth estimate $\widehat{f}$. The basis expansion~\eqref{BSplineExpansion} allows us to express the nonparametric regression model~\eqref{NonparametricRegressionModel} in the form of a multiple linear regression model
\begin{align*}
    y=Bb+\epsilon,\ \epsilon\sim N_n(0,\sigma^2 I_n),
\end{align*}
where $y=(y_1,\dots,y_n)^T$ is the vector of observations and $B$ is the $n\times D$ tensor product B-spline design matrix. The overall dimension $D=\prod_{j=1}^p d_j$ of the tensor product spline space can be very large, e.g. for a five-dimensional smooth ($p=5$) and ten-dimensional marginal bases ($d=10$) we already have $D=10^5=100,000$.

\subsection{Anisotropic Smoothness Prior}
To obtain a smooth estimate $\widehat{f}$, we introduce a vector $\tau^2=(\tau_1^2,\dots,\tau_p^2)^T$ of positive smoothing variances and endow the tensor product B-spline coefficients with the partially improper Gaussian prior
\begin{align}\label{ConditionalPriorB}
    p(b\mid \tau^2) \propto Det(K(\tau^2))^{1/2} \ \exp\left(-\dfrac{1}{2} b^TK(\tau^2)b\right),\ b\in\dsR^D.
\end{align}
Thereby, $Det$ is the pseudo-determinant (a.k.a.~generalized determinant) which is defined as the product of nonzero eigenvalues \citep{Kni2014} and $K(\tau^2)$ is the overall roughness penalty matrix of the form
\begin{align}\label{OverallPenaltyMatrix}
    K(\tau^2)= \dfrac{K_1}{\tau_1^2}+\dots +\dfrac{K_p}{\tau^2_p},
\end{align}
with $K_j= I_{d_1}\otimes \dots \otimes I_{d_{j-1}}\otimes \widetilde{K}_j \otimes I_{d_{j+1}}\otimes \dots \otimes I_{d_p},\ j=1,\dots,p$ \citep[cf.][]{EilMar2003,Woo2006b}. Furthermore, $\otimes$ denotes the Kronecker product and $\widetilde{K}_j$ is the usual $d_j\times d_j$ P-spline second order differences penalty matrix corresponding to the $j$-th marginal B-spline basis \citep[see][for further details]{EilMar1996}. 

By using an entire vector $\tau^2=(\tau_1^2,\dots,\tau_p^2)^T$ instead of a single smoothing variance in~\eqref{ConditionalPriorB} and~\eqref{OverallPenaltyMatrix}, we allow for a different amount of smoothing for each coordinate $x_j,\ j=1,\dots,p$. This is crucial to achieve satisfactory estimation accuracy. Motivated by their popularity in the context of additive models, we consider two different choices for the prior $p(\tau^2)$ of the smoothing variances. We consider Inverse Gamma priors \citep[cf.][]{FahKneLan2004}
\begin{align}\label{InverseGammaPrior}
    \tau_j^2\overset{ind}{\sim} IG(\alpha_j,\beta_j),\ j=1,\dots, p,
\end{align}
as well as Weibull priors with shape $1/2$ \citep[cf.][]{KleKne2016}, i.e.
\begin{align}\label{WeibullPrior}
    \tau_j^2\overset{ind}{\sim} Weibull(1/2,\lambda_j),\ j=1,\dots, p.
\end{align}

To complete the prior specification, we place the Jeffreys' prior on the unknown residual variance $\sigma^2>0$, i.e. we use $p(\sigma^2) \propto 1/\sigma^2$. 

\section{Efficient Posterior Sampling}\label{sec:EfficientMCMC}
In this section we derive a highly efficient MCMC sampler for the Bayesian anisotropic P-spline model. By Bayes' rule, the joint posterior is proportional to
\begin{align*}
    p(b,\tau^2,\sigma^2\mid y) \propto p(y\mid b,\sigma^2)\ p(b\mid \tau^2)\ p(\tau^2)\ p(\sigma^2),\ (b,\tau^2,\sigma^2)\in \dsR^D\times (0,\infty)^p\times (0,\infty),
\end{align*}
which does not correspond to a known probability distribution. Therefore, we use MCMC methods to generate a sample from the joint posterior. In what follows, we first address the MCMC update of the tensor product B-spline coefficients and the residual variance, then we address the MCMC update of the vector of smoothing variances.

\subsection{Updating the Regression Coefficients and the Residual Variance}
The full conditional posterior of the tensor product B-spline coefficients is a multivariate Gaussian distribution 
\begin{align}\label{FullConditionalPosteriorB}
    b\mid \tau^2,\sigma^2,y \sim N_D\left((B^TB/\sigma^2+K(\tau^2))^{-1}B^Ty/\sigma^2,(B^TB/\sigma^2+K(\tau^2))^{-1}\right),
\end{align}
which is straightforward to sample from. Moreover, there are several strategies to increase computational efficiency: One can e.g.~use a sparse Cholesky decomposition of the precision matrix $B^TB/\sigma^2+K(\tau^2)$ or one can use blockwise updates if the dimension $D$ is very large (see Appendix Section~A for further details). The full conditional posterior of the residual variance is Inverse Gamma
\begin{align}\label{FCPsigma2}
    \sigma^2 \mid b,y\sim IG(n/2,\|y-Bb\|_2^2/2),
\end{align}
which is very straightforward to sample from.

\subsection{Updating the Smoothing Variances}
The full conditional posterior of the vector of smoothing variances $\tau^2$ is proportional to 
\begin{align}\label{FCPtau2}
    p(\tau^2\mid b) \propto Det(K(\tau^2))^{1/2} \ \exp\left(-\dfrac{1}{2} b^TK(\tau^2)b\right)\ p(\tau^2),\ \tau^2 \in (0,\infty)^p,
\end{align}
which does not correspond to a known probability distribution, irrespective of the prior $p(\tau^2)$. This is because of the pseudo-determinant $Det(K(\tau^2))$ and implies that we cannot use Gibbs steps for the vector of smoothing variances $\tau^2$. Moreover, deriving efficient MH updates for $\tau^2$ is challenging because any MH update necessarily involves the repeated computation and evaluation of the pseudo-determinant 
\begin{align*}
    Det(K((\tau^2)^\ast))
\end{align*}
at a proposed value $(\tau^2)^\ast$, which generally has a high computational burden: The most obvious approach to compute this pseudo-determinant is to perform an eigendecomposition of the $D\times D$ penalty matrix $K((\tau^2)^\ast)$. However, despite the sparsity of the penalty matrix, the eigendecomposition has computational complexity $\mathcal{O}(D^3)$. Therefore, the update of $\tau^2\in(0,\infty)^p$ is in fact much more expensive than the update of $b\in\dsR^D$ even though $p\ll D$. To address this challenge, we exploit the following simple yet previously unrecognized expressions for the penalty matrix and its pseudo-determinant.

\subsubsection{Simple Expressions for the Penalty Matrix and its Determinant}
\begin{theorem}[Penalty matrix decomposition]\label{Theo:KeyResult} 
Let $\widetilde{K}_j=\widetilde{Q}_j\widetilde{\Gamma}_j\widetilde{Q}^T_j,\ j=1,\dots,p,$ be eigendecompositions of the marginal penalty matrices. Let $Q=\widetilde{Q}_1\otimes\dots \otimes \widetilde{Q}_p$ and $\Gamma_j=I_{d_1}\otimes \dots \otimes I_{d_{j-1}}\otimes \widetilde{\Gamma}_j \otimes I_{d_{j+1}}\otimes \dots \otimes I_{d_p},\ j=1,\dots,p$. Then, for all $\tau^2\in(0,\infty)^p$ it holds:
\begin{align}\label{SpecialRepresentation}
    K(\tau^2)=Q \left(\dfrac{\Gamma_1}{\tau_1^2}+\dots +\dfrac{\Gamma_p}{\tau_p^2}\right) Q^T.
\end{align}
\end{theorem}

Theorem~\ref{Theo:KeyResult} follows from the definition of the overall roughness penalty matrix~\eqref{OverallPenaltyMatrix} and the properties of the Kronecker product. A proof is provided in Section B of the Appendix. Theorem~\ref{Theo:KeyResult} implies the following convenient expression for the log-pseudo-determinant. 

\begin{corollary}[Log-determinant]\label{Cor:logDeterminant} Let $\gamma_{j,l},\ j=1,\dots,p,\ l=1,\dots,D,$ denote the diagonal entries of the $\Gamma_j,\ j=1,\dots,p,$ and let the set $\mathcal{D}^+\subseteq\{1,\dots,D\}$ contain those indices where at least one of the $\Gamma_j,\ j=1,\dots,p,$ has a positive diagonal entry, i.e. $l\in \mathcal{D}^+ \iff \exists\ j\in\{1,\dots,p\}: \gamma_{j,l}>0$. Then, for all $\tau^2\in(0,\infty)^p$ it holds:
\begin{align}\label{logDeterminant}
    \log Det(K(\tau^2))=\log Det\left(\dfrac{\Gamma_1}{\tau_1^2}+\dots +\dfrac{\Gamma_p}{\tau_p^2}\right)=\sum_{l\in \mathcal{D}^+} \log \left(\dfrac{\gamma_{1,l}}{\tau_1^2}+\dots + \dfrac{\gamma_{p,l}}{\tau_p^2}\right).
\end{align}
\end{corollary}

A proof of Corollary~\ref{Cor:logDeterminant} is provided in Section~B of the Appendix. Corollary~\ref{Cor:logDeterminant} reduces the numerical cost of the computation of $Det(K(\tau^2))$ from cubic complexity $\mathcal{O}(D^3)$ to linear complexity $\mathcal{O}(D)$. Therefore, the evaluation of the full conditional posterior~\eqref{FCPtau2} becomes much cheaper and efficient MH updates for $\tau^2$ become feasible. 

\begin{remark}
The penalty matrix decomposition appears in various different forms for the special cases $p=2$ or $p=3$ in the penalized splines literature \citep[see, e.g.,][]{LeeDur2011,RodLeeKne2015,KneKleLan2019}. However, it seems that the generality and in particular the practical usefulness of this result for the fully Bayesian P-splines approach have not been recognized so far. 
\end{remark}

\subsubsection{Taylored MH Updates} \label{Sec:MHUpdates}
The main idea of our new approach is to exploit the simple expression~\eqref{logDeterminant} to derive efficient and adaptive MH updates for the smoothing parameters. The basic idea of these updates is to approximate the target density locally by a (multivariate) Gaussian density. These updates are known as Taylored or iteratively weighted least squares (IWLS) updates in the literature \citep[cf.][]{GewTan2003,KleKne2016}. In the present context, the target density is the full conditional posterior of the log-smoothing variances. We work with the log-smoothing variances $\rho_j=\log (\tau_j^2),\ j=1,\dots,p,$ as these are unconstrained. 

By the density transformation formula and~\eqref{FCPtau2}, the full conditional posterior of the log-smoothing variances $\rho=(\rho_1,\dots,\rho_p)\in\dsR^p$ is proportional to
\begin{align*}
    p(\rho\mid b) \propto Det(K(e^\rho))^{1/2} \ \exp\left(-\dfrac{1}{2} b^TK(e^\rho)b\right)\  q(\rho),\ 
\end{align*}
where $q(\rho)\propto p(e^\rho)\ \prod_{j=1}^p e^{\rho_j}$ is a kernel of the prior of the log-smoothing variances $\rho$ and $e^\rho=(e^{\rho_1},\dots,e^{\rho_p})^T$. By Corollary~\ref{Cor:logDeterminant}, the log-full conditional posterior of $\rho$ is (up to an irrelevant additive constant) equal to
\begin{align}\label{logFCPrho}
   \log p(\rho\mid b)=\dfrac{1}{2}\sum_{l\in D^+} \log \left(\gamma_{1,l}e^{-\rho_1}+\dots + {\gamma_{p,l}}e^{-\rho_p}\right)-\dfrac{1}{2} b^TK(e^\rho)b + \log q(\rho).
\end{align}
 Following \cite{GewTan2003,KleKne2016} our adaptive MH proposal $\rho^\ast$ relies on the gradient vector $u(\rho)$ and the Hessian matrix $H(\rho)$ of the log-full conditional posterior~\eqref{logFCPrho}. In the subsequent proposition we state the corresponding first and second order partial derivatives.

\begin{proposition}[Partial derivatives]\label{Prop:PartialDerivatives} For $\rho\in\dsR^p$ and $j_0, k_0\in\{1,\dots,p\}$ it holds
\begin{align*}
    \partial_{{j_0}} \log p(\rho\mid b)&=-\dfrac{1}{2} \sum_{l\in D^+} \dfrac{\gamma_{j_0,l}e^{-\rho_{j_0}}}{\gamma_{1,l}e^{-\rho_1}+\dots + {\gamma_{p,l}}e^{-\rho_p}} +\dfrac{1}{2} b^TK_{j_0}b\ e^{-\rho_{j_0}} +\partial_{j_0}\log q(\rho),\\
    \partial^2_{{j_0}}\log p(\rho\mid b)&=-\dfrac{1}{2} \sum_{l\in D^+} \left\lbrace \left(\dfrac{\gamma_{j_0}e^{-\rho_{j_0}}}{\gamma_{1,l}e^{-\rho_1}+\dots + {\gamma_{p,l}}e^{-\rho_p}}\right)^2-\dfrac{\gamma_{j_0}e^{-\rho_{j_0}}}{\gamma_{1,l}e^{-\rho_1}+\dots + {\gamma_{p,l}}e^{-\rho_p}}\right\rbrace\\
    &\quad
   -\dfrac{1}{2} b^TK_{j_0}b\ e^{-\rho_{j_0}} +\partial^2_{j_0}\log q(\rho),\\
     \partial_{{j_0}}\partial_{{k_0}}\log p(\rho\mid b)&= -\dfrac{1}{2} \sum_{l\in D^+} \dfrac{\gamma_{j_0,l}e^{-\rho_{j_0}}\gamma_{k_0,l}e^{-\rho_{k_0}}}{(\gamma_{1,l}e^{-\rho_1}+\dots + {\gamma_{p,l}}e^{-\rho_p})^2}+\partial_{{j_0}}\partial_{{k_0}}\log q(\rho),\quad \mbox{ if }  j_0\neq k_0. 
\end{align*}
\end{proposition}

While the expressions in Proposition~\ref{Prop:PartialDerivatives} appear complicated at first sight, it is important to realize that they can be evaluated very efficiently. 
Next we explain how these expressions can be used to generate Taylored MH updates for the log-smoothing variances $\rho$. Given that the current B-spline coefficients are $b^{(t+1)}\in\dsR^D$ and the current position of the log-smoothing variances is $\rho^{(t)}\in\dsR^p$, a single MH step for $\rho$ goes as follows:

\begin{enumerate}
    \item Generate the proposal $\rho^\ast$ from the $p$-variate Gaussian distribution
    \begin{align*}
     N_p(\mu^{(t)},\Sigma^{(t)})
\end{align*}
with mean vector $\mu^{(t)}=\rho^{(t)}-{H}^{-1}(\rho^{(t)})u(\rho^{(t})$ and covariance matrix $\Sigma^{(t)}=-{H}^{-1}(\rho^{(t)})$. 
\item Compute the MH acceptance probability
\begin{align*}
    \alpha^\ast=\text{min}\left(1,\dfrac{p(\rho^\ast\mid b^{(t+1)})}{p(\rho^{(t)}\mid b^{(t+1)} )}\dfrac{N_p(\rho^{(t)};\mu^\ast,\Sigma^\ast)}{N_p(\rho^\ast;\mu^{(t)},\Sigma^{(t)})}\right),
\end{align*}
where $N_p(\rho;\mu,\Sigma)$ denotes the density of a $p$-variate Gaussian distribution with mean vector $\mu$ and covariance matrix $\Sigma$ evaluated at $\rho$. Moreover, $\mu^\ast=\rho^\ast-H^{-1}(\rho^\ast)u(\rho^\ast)$ and $\Sigma^\ast=-H^{-1}(\rho^\ast)$.
\item Set $\rho^{(t+1)}=\rho^\ast$ with probability $\alpha^\ast$ and $\rho^{(t+1)}=\rho^{(t)}$ with probability $1-\alpha^\ast$.
\end{enumerate}
To implement the Taylored MH updates for the two priors~\eqref{InverseGammaPrior} and~\eqref{WeibullPrior} we need the log-kernel in terms of the log-smoothing variances, i.e.~$\log q(\rho)$, as well as the corresponding first and second order partial derivatives (see Section~E of the Appendix for details). Combining the Taylored MH steps for $\rho$ with Gibbs steps for $b$ and $\sigma^2$ from~\eqref{FullConditionalPosteriorB} and~\eqref{FCPsigma2}, respectively, we obtain a MCMC sample from the joint posterior $(b,\rho,\sigma^2)\mid y$. This sample can then be used to make inference about the unknown function $f$ in standard Bayesian fashion. 

For the initialization of our MCMC sampler we replace the first $100$ MH steps for $\rho$ by Newton-Raphson steps and, in addition to that, we standardize $y$ to have mean zero and unit variance. Next we discuss another important detail that guarantees the numerical stability of our algorithm.

\subsubsection{Hessian Modification}\label{HessianMod}

For the MH update of  the previous section to be well defined we need both Hessians $H(\rho^{(t)})$ and $H(\rho^{\ast})$ to be negative definite (otherwise the MH acceptance probability $\alpha^\ast$ is not well-defined). To ensure that this is the case, we follow Section 3.4 of \cite{NocWri2006} in the context of Newton's method and modify the eigenvalues of the Hessian, if they are not already sufficiently small. To this end, we replace the eigenvalues $\lambda_j,\ j=1,\dots,p,$ of the Hessian matrix $H$ by 
 \begin{align}\label{HessianModification}
 \widetilde{\lambda}_j=\text{min}(\lambda_j,-\delta),\ j=1,\dots,p,    
 \end{align}
  where $\delta>0$ is a fixed positive constant that is chosen by the user. Denoting the modified Hessian matrix by $\widetilde{H}$, we thus use the matrices $-\widetilde{H}(\rho^{(t)})^{-1}$ and $-\widetilde{H}(\rho^{\ast})^{-1}$ in our MCMC scheme. By default we use $\delta=1/\pi$ for the threshold, which ensures that the Hessians are negative definite and, in addition to that, limits the maximal step size to a reasonable range. Similar modifications of the Hessian matrix are also common for REML based approaches \citep[see, e.g., Section 3 of][]{Woo2011}.
 
\section{Main Effects and Interactions}\label{sec:MainEffectsAndInteractions}
In this section we develop general formulas that allow for efficient derivation of lower-dimensional effects such as one-dimensional main effects and two-dimensional interactions for an arbitrary dimensional tensor product smooth. To the best of our knowledge, these formulas have not been established in the literature before.

The approach introduced in the previous sections allows for efficient Bayesian estimation of a tensor product smooth $\widehat{f}=\widehat{f}(x_1,\dots,x_p)$ of moderate dimension $p\in\{2,3,4,\dots\}$. An important question in practice is how such a smooth can be interpreted if $p>2$. This is not completely obvious because for $p>2$ the function graph cannot be plotted anymore. One straightforward option facilitating interpretation are slice plots. Thereby, we fix some of the coordinates and regard $\widehat{f}$ as a function of the remaining coordinates only.
Another option that is closely related to functional ANOVA decompositions \citep[see, e.g.,][]{LeeDur2011,Gu2013} are plots of the one-dimensional main effects and the two-dimensional interactions. As one can find different definitions of these notions in the literature \citep[cf.][]{Sto1994,Hoo2007,Gu2013}, we start with a precise definition to clarify what we refer to.

\begin{definition}
Let $f=f(x_1,\dots,x_p)$ be a tensor product spline. Then we define the main effect of $x_j$ as the function that is obtained by integrating all other coordinates out, i.e.~the $j$-th main effect is defined as
\begin{align*}
    x_j\mapsto \int_{[0,1]^{p-1}} f(x_1,\dots,x_p)dx_{-j},
\end{align*}
where $x_{-j}$ denotes the vector $x=(x_1,\dots,x_p)$ without the $j$-th component. Similarly, the two-dimensional interaction of $x_j$ and $x_k$ is defined as the function
\begin{align*}
     (x_j,x_k)\mapsto \int_{[0,1]^{p-2}} f(x_1,\dots,x_p)d x_{-{(j,k)}},\
\end{align*}
where $x_{-{(j,k)}}$ denotes $x=(x_1,\dots,x_p)$ without the $j$-th and $k$-th component.
\end{definition}

Next, we show how the main effects and two-dimensional interactions can be derived very efficiently for a $p$-dimensional tensor product smooth. Proposition~\ref{Prop:IntegrateOut} is the key result.
 
\begin{proposition}\label{Prop:IntegrateOut} Let $f=f(x_1,\dots,x_p)$ be a tensor product spline. Then the function
\begin{align}\label{NewFunc}
    x_{-j}\mapsto \int_{[0,1]} f(x_1,\dots,x_p) dx_j 
\end{align}
is a tensor product spline in the smaller tensor product space that is obtained when omitting the $j$-th marginal spline basis for the tensor product. If $b\in\dsR^D$ are the coefficients of $f$ with respect to the tensor product B-splines, then 
\begin{align*}
    (I_{d_1} \otimes \dots \otimes I_{d_{j-1}} \otimes A_j \otimes I_{d_{j+1}}\otimes \dots \otimes  I_{d_p})\ b 
\end{align*}
are the coefficients of~\eqref{NewFunc} with respect to the tensor product B-splines in the smaller tensor product spline space. Thereby, $A_j$ is a $1\times d_j$ matrix that contains the averages of the $j$-th marginal B-spline basis, i.e.
    \begin{align*}
        A_j=\left(\int_{[0,1]} B_1(x_j)dx_j,\dots, \int_{[0,1]} B_{d_j}(x_j)dx_j\right).
    \end{align*}
\end{proposition}

\begin{corollary}\label{Cor:MainInteractions}
The $j$-th main effect is in the span of the $j$-th marginal B-spline basis and the corresponding coefficients are
\begin{align}\label{formulaMainEffect}
    (A_1 \otimes \dots \otimes A_{j-1}\otimes I_{d_j} \otimes A_{j+1}\otimes \dots \otimes A_p)\ b.
\end{align}
The two-dimensional interaction of $x_j$ and $x_k$ is in the tensor product space spanned by the $j$-th and $k$-th marginal B-spline bases and the corresponding coefficients are
\begin{align}\label{formulaInteraction}
    (A_1 \otimes \dots \otimes A_{j-1} \otimes I_{d_j} \otimes A_{j+1} \otimes \dots \otimes  A_{k-1} \otimes I_{d_k} \otimes A_{k+1} \otimes \dots \otimes A_p)\ b.
\end{align}
\end{corollary}

\begin{example} To give a concrete example we consider a three-dimensional tensor product spline ${f}(x_1,x_2,x_3)$ and apply Corollary~\ref{Cor:MainInteractions}. The results are summarized in Table~\ref{tab:ExampleLowerDimEffects}.

\begin{table}[H]
    \centering
    \begin{tabular}{c|c|c}
         Effect  & Basis & Coefficients\\ 
         \hline \hline
          Main effect of $x_1$ & $\{B_1(x_1),\dots,B_{d_1}(x_1)\}$ &  $(I_{d_1}\otimes A_2\otimes A_3)\ {b}$\\
          Main effect of $x_2$ & $\{B_1(x_2),\dots,B_{d_2}(x_2)\}$ &   $(A_1 \otimes I_{d_2}\otimes A_3)\ {b}$\\
          Main effect of $x_3$ & $\{ B_1(x_3),\dots,B_{d_3}(x_3)\}$ &   $(A_1\otimes A_2\otimes I_{d_3})\ {b}$\\
          Interaction of $x_1$ and $x_2$ & $\{B_1(x_1),\dots,B_{d_1}(x_1)\}\otimes \{B_1(x_2),\dots,B_{d_2}(x_2)\} $ &   $(I_{d_1}\otimes I_{d_2}\otimes A_3)\ {b}$\\
         Interaction of $x_1$ and $x_3$ & $\{B_1(x_1),\dots,B_{d_1}(x_1)\}\otimes \{B_1(x_3),\dots,B_{d_3}(x_3)\}$ &  $(I_{d_1}\otimes A_2\otimes I_{d_3})\ {b}$\\
          Interaction of $x_2$ and $x_3$ & $\{B_1(x_2),\dots,B_{d_2}(x_2)\}\otimes \{B_1(x_3),\dots,B_{d_3}(x_3)\}$ &   $(A_1\otimes I_{d_2}\otimes I_{d_3})\ {b}$
    \end{tabular}
    \caption{Spline basis and coefficients of the one-dimensional main effects and the two-dimensional interactions for a three-dimensional tensor product spline $f(x_1,x_2,x_3).$}
    \label{tab:ExampleLowerDimEffects}
\end{table}
\end{example}

The results of this section show that given an estimate $\widehat{f}=\sum_{j=1}^D B_j\widehat{b}_j$ it is straightforward to derive estimates for the main effects and the two-dimensional interactions. To this end, we simply apply formulas~\eqref{formulaMainEffect} and~\eqref{formulaInteraction} to the estimated coefficient vector $\widehat{b}$. Therefore, we only need the averages $A_j,\ j=1,\dots,p,$ of the marginal B-spline bases (cf. Proposition~\ref{Prop:IntegrateOut}). Formulas for the corresponding one-dimensional integrals are readily available in the literature~\citep[see, e.g.,][page 128]{Boo2001} and have for example been implemented in the \texttt{R} package \texttt{IntegrateBs} \citep{Bai2016}.

\begin{remark}
We can also derive credible intervals for the one-dimensional main effects and the two-dimensional interactions by simply applying formulas~\eqref{formulaMainEffect} and~\eqref{formulaInteraction} to the entire MCMC sample of tensor product B-spline coefficients $b^{(t)},\ t=1,\dots,T$. In conjunction with the corresponding design matrices we then obtain a whole sample of main effects and two-dimensional interactions from which we can derive pointwise credible intervals. Using the approach of \cite{KriKneCla2010} we can also derive simultaneous credible intervals. This approach is e.g.~implemented in the \texttt{R} package \texttt{acid} \citep{Soh2016}.
\end{remark}

\section{Empirical Evidence}\label{sec:EmpiricalEvidence}

In this section we provide empirical evidence for our new approach. First we conduct a simulation study, then we consider a real data example. All computations were conducted in \texttt{R} \citep{RCore22} on a regular desktop PC with 3.5 GHz and 32 GB RAM.

\subsection{Simulation Study}\label{sec:SimulationStudy}
Our simulation study is divided into two parts.

\begin{enumerate}[a)]
    \item The first part focuses on computational efficiency. Specifically, we aim to answer: \begin{itemize}
    \item For which combinations of the dimensions $p\in\{2,3,4,5\}$ and $d\in\{5,10\}$ is the runtime of our MCMC sampler acceptable? Recall that $p$ is the dimension of the tensor product smooth and $d$ is the dimension of the marginal B-spline bases.
    \item What about the runtime of competitors such as \texttt{bamlss}, \texttt{BayesX}, \texttt{jagam} or \texttt{rstanarm}?
\end{itemize}
    \item The second part focuses on estimation accuracy. Specifically, we aim to answer:
    \begin{itemize}
        \item How do we compare with respect to competitors such as \texttt{bamlss}, \texttt{BayesX}, \texttt{jagam} or \texttt{rstanarm} in terms of MSE?
        \item What if the test function is isotropic? What if the test function is anisotropic? 
    \end{itemize}
\end{enumerate}

\subsubsection*{Part a) Computational Efficiency (runtime)}
We consider the isotropic test function
$
    f_1(x)=\sin(2\pi \|x\|_2),\ x\in[0,1]^p.
$
We use a sample size of $n=10^4$ and the residual variance $\sigma^2$ is set to $(1/2)^2$. The design points $x_1,\dots,x_n$ are sampled iid and uniformly on the $p$-dimensional unit cube $[0,1]^p$. We increase the dimension of the domain $p\in\{2,3,4,5\}$ and record the time needed to generate $1,200$ MCMC samples for the following five methods:
\begin{itemize}
    \item new-WB: Our new approach with iid unit rate Weibull priors for the smoothing variances, i.e. $\tau_j^2\overset{iid}{\sim} Weibull(1/2,1),\ j=1,\dots,p$.  
    \item bamlss: The function \texttt{bamlss} in the \texttt{R} package \texttt{bamlss} with sampler \texttt{sam\_GMCMC}. 
\item BayesX: The function \texttt{bamlss} in the \texttt{R} package \texttt{bamlss} with sampler \texttt{sam\_BayesX}. 
\item jagam: The function \texttt{jagam} in the \texttt{R} package \texttt{mgcv}. 
\item rstanarm: The function \texttt{stan\_gamm4} in the \texttt{R} package \texttt{rstanarm}. 
\end{itemize}
 For all five methods we consider either $d=5$ or $d=10$ for the dimensions of the marginal B-spline bases. Table~\ref{tab:Runtime} shows the runtime for each of the competitors in minutes.
 
\begin{table}[H]
\centering
\begin{tabular}{l|l|rrrr}

 & Method & $p=2$ & $p=3$ & $p=4$ & $p=5$ \\ 
 \hline \hline
$d=5$ & new-WB & 0.4 & \textbf{0.7} & \textbf{2.7} & \textbf{28.70} \\ 
 &  bamlss & 1.2 & 6.3 & 151.3 & $>$ 600 \\ 
 &   BayesX & 0.57 & 1.51 & $>$ 600 &  \\ 
 & jagam & \textbf{0.22} & 4.23 & 346.83 & $>$ 600 \\ 
 &  rstanarm & 0.37 & 2.82 & 72.03 & $> 600$ \\ 
  \hline
$d=10$ &  new-WB & 0.3 & \textbf{2.7} & \textbf{26.32} & \textbf{259.53} \\ 
 & bamlss & 3.15 & $>$ 600 & $>$ 600 & $>$ 600  \\ 
 & BayesX & 0.58 & 7.97 & $>$ 600 &  \\ 
 & jagam & 2.42 & $>$ 600 & $>$ 600 & $>$ 600 \\ 
 & rstanarm & \textbf{0.21} & 24.62 & $>$ 600 & $>$ 600 \\ 
\end{tabular}\caption{Runtime in minutes to generate 1,200 MCMC samples. 
The rows show the different methods, while across the columns, the dimension $p$ of the tensor product smooth increases. The MCMC sampling was interrupted after 10 hours (marked with $>$ 600). Two values are missing in the rightmost column because \texttt{BayesX} currently does not support five-dimensional tensor product smooths.} 
    \label{tab:Runtime}
\end{table}

\subsubsection*{Conclusions}
\begin{itemize}
    \item For a two-dimensional tensor product smooth $(p=2)$ our new approach is a few seconds slower than some of the competitors. 
    \item However, for dimension $p=3$ or higher, our new approach is magnitudes faster than previous fully Bayesian approaches allowing for anisotropic multidimensional smoothing. This is true for five-dimensional marginal bases $(d=5)$ and in particular for ten-dimensional marginal bases $(d=10)$.
\end{itemize}

\subsubsection*{Part b) Estimation Accuracy (MSE)}
\begin{figure}[htbp]
    \centering
    \includegraphics[width=0.9\textwidth]{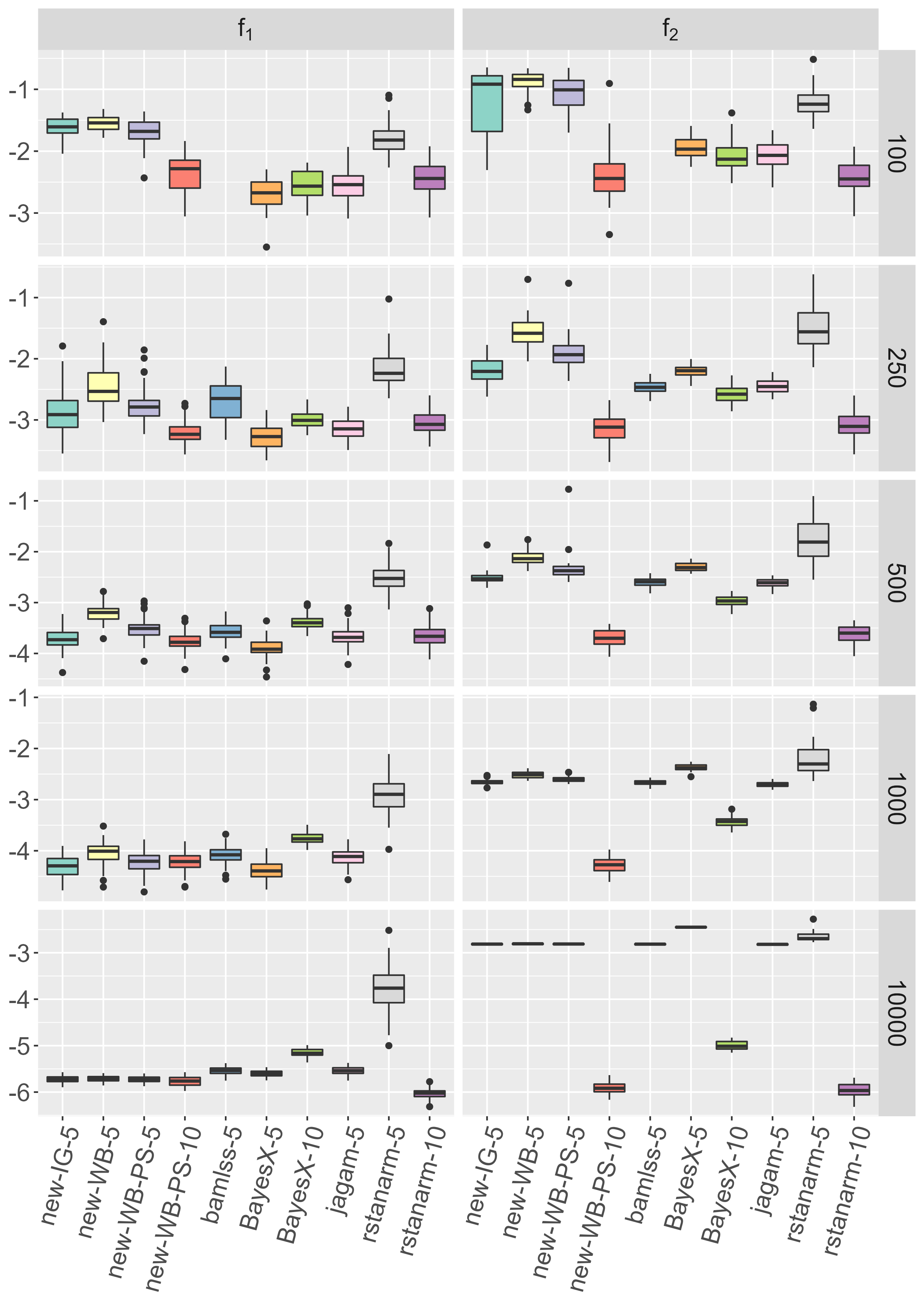}
    \caption{Shown are the log MSEs for the different methods. The five plots on the left show the results for the isotropic test function $f_1$, the five plots on the right show the results for the anisotropic test function $f_2$. The rows show the results for the different sample sizes $n\in\{100,250,500,10^3,10^4\}$. The suffix in the labels indicates the dimension $d$ of the marginal B-spline bases, e.g. $d=5$ for rstanarm-5 and $d=10$ for rstanarm-10.}
    \label{fig:logMSE}
\end{figure}

Next we fix the dimension of the domain $p=3$ and thus only consider three-dimensional tensor product smooths. In addition to the isotropic test function $f_1$ we consider the anisotropic test function $f_2(x)=\sin\left(2\pi \sqrt{3x_1^2+x_2^2+x_3^2/3}\right),\ x\in[0,1]^3.$ We vary the sample size $n\in\{100,250,500,10^3,10^4\}$ and compute the $
MSE=\dfrac{1}{n}\sum_{i=1}^n (\widehat{f}(x_i)-f(x_i))^2.
$
The design points $x_1,\dots,x_n$ are sampled iid and uniformly on the three-dimensional unit cube $[0,1]^3$ and the residual variance $\sigma^2$ is set to $(1/2)^2$ as before. We use $1,200$ MCMC iterations and discard the first $200$ as burnin. We consider our new approach with four different parameter settings:
\begin{enumerate}[i)]
    \item Inverse Gamma priors $\tau_j^2\overset{iid}{\sim}IG(0.001,0.001)$ and $d=5$
        \item Weibull priors $\tau_j^2\overset{iid}{\sim}Weibull(1/2,1)$ and $d=5$
        \item Weibull priors $\tau_j^2\overset{iid}{\sim}Weibull(1/2,\lambda)$ with $\lambda$ determined via prior scaling and $d=5$. The key idea of the prior scaling approach is to set $\lambda$ such that prior function draws have a reasonable scale (further details are provided in  Appendix Section~D). 
        \item Weibull priors $\tau_j^2\overset{iid}{\sim}Weibull(1/2,\lambda)$ with $\lambda$ determined via prior scaling and $d=10$ dimensional marginal bases. 
    \end{enumerate}
As further competitors we consider bamlss and jagam with $d=5$ as well as BayesX and rstanarm with $d=5$ or $d=10$. We do not include $d=10$ for {bamlss} and {jagam} because of the excessive runtime established before
(see Table~\ref{tab:Runtime}). Figure~\ref{fig:logMSE} shows boxplots of the log MSE based on $R=50$ replicates for each configuration of $f\in \{f_1,f_2\}$ and $n\in\{100,250,500,10^3,10^4\}$. 

\subsubsection*{Conclusions}
\begin{itemize}
    \item Altogether, parameter setting iv) with label ``new-WB-PS-10'' in Figure~\ref{fig:logMSE} works best for our new approach, i.e. Weibull priors $\tau_j^2\overset{iid}{\sim}Weibull(1/2,\lambda)$ with $\lambda$ determined via prior scaling and $d=10$ dimensional marginal bases yield the best performance. Therefore, we opt for this setting as default. With this setting, we are slightly worse than some of the competitors for the isotropic test function $f_1$. However, we outperform all of the competitors for the anisotropic test function $f_2$. The only method that can keep up is rstanarm-10. However, we found rstanarm-10 to be unreliable in the sense that the MCMC sampler typically got stuck for the sample size $n=10^3$. This is also the reason why the corresponding boxes in the fourth row of Figure~\ref{fig:logMSE} are missing for rstanarm-10.
    \item In contrast to that, we did not encounter any numerical issues for our new approach. Interestingly, the modification of the Hessian (see Section~\ref{HessianMod}) was only necessary for the Inverse Gamma prior $\tau_j^2\overset{iid}{\sim}IG(0.001,0.001)$ but not for the Weibull prior. More specifically, the eigenvalue modification allowed us to avoid numerical issues such as an indefinite Hessian in about 10\% of the runs for the Inverse Gamma prior. For the Weibull prior, however, the modification was never exerted. This can be explained by the much lighter tails of the Weibull prior which ensure that the parameters stay within a reasonable range during MCMC sampling. The key message is that the Weibull prior offers better numerical stability compared to the Inverse Gamma prior. This finding is in line with the observations of others \cite[see, e.g.,][]{GhoLiMit2018}.
\end{itemize}

\paragraph{Overall summary.}
 In summary, the simulation study shows that our new approach is much faster than previous Bayesian approaches allowing for anisotropic multidimensional smoothing. Moreover, the new approach is numerically stable and performs equally well or even better in terms of MSE. 
 
 \newpage
\subsection{Real data example}~\label{sec:RealDataExample}
In this section we apply our new approach to analyze a publicly available temperature data set. The data set comprises $n=12,672$ records of the monthly average temperature from January 2000 to December 2010 for $96$ measurement locations across the USA. The temperature data set is part of a large climate data base that was compiled by the Berkeley Earth project (\url{www.berkeleyearth.org}). Further information about the data set and our preprocessing steps are provided in Section F of the Appendix. 
Figure~\ref{fig:DataSet} visualizes the temperature data set.

\begin{figure}[htbp]
    \centering
    \includegraphics[width=\textwidth,height=2.25in]{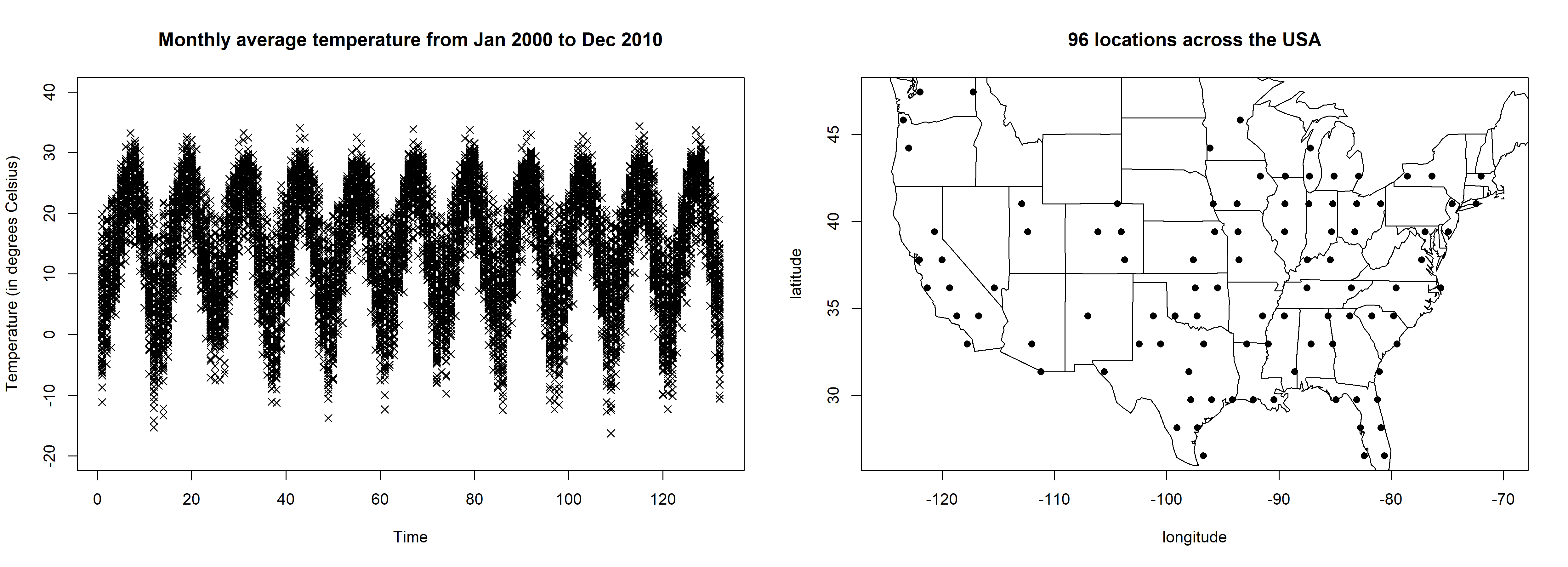}
    \caption{Left: Monthly average temperature from January 2000 to December 2010 for $96$ measurement locations across the USA. Right: 96 measurement locations across the USA.} 
    \label{fig:DataSet}
\end{figure}

To gain insights into the spatio-temporal temperature dynamics, we consider the spatio-temporal model
\begin{align}\label{SpaceTimeModel}
    Temperature=f(time,longitude,latitude)+\epsilon,\ \epsilon\sim N(0,\sigma^2).
\end{align}
We model $f$ as a three-dimensional tensor product smooth using the following parameters: We use $(d_1,d_2,d_3)=(40,10,10)$ for the dimensions of the marginal B-spline bases. The overall dimension of the tensor product spline space is thus $D=40\times 10\times 10=4,000$. For the smoothing variances we use independent Weibull priors $\tau_j^2\overset{iid}{\sim}Weibull(1/2,\lambda),\ j=1,2,3,$ where $\tau^2_1$ refers to time, $\tau^2_2$ refers to longitude and $\tau^2_3$ refers to latitude. The rate parameter $\lambda\approx 38.37$ was determined via prior scaling (see Appendix Section D for details).

\begin{figure}[htbp]
    \centering
    \includegraphics[width=\textwidth]{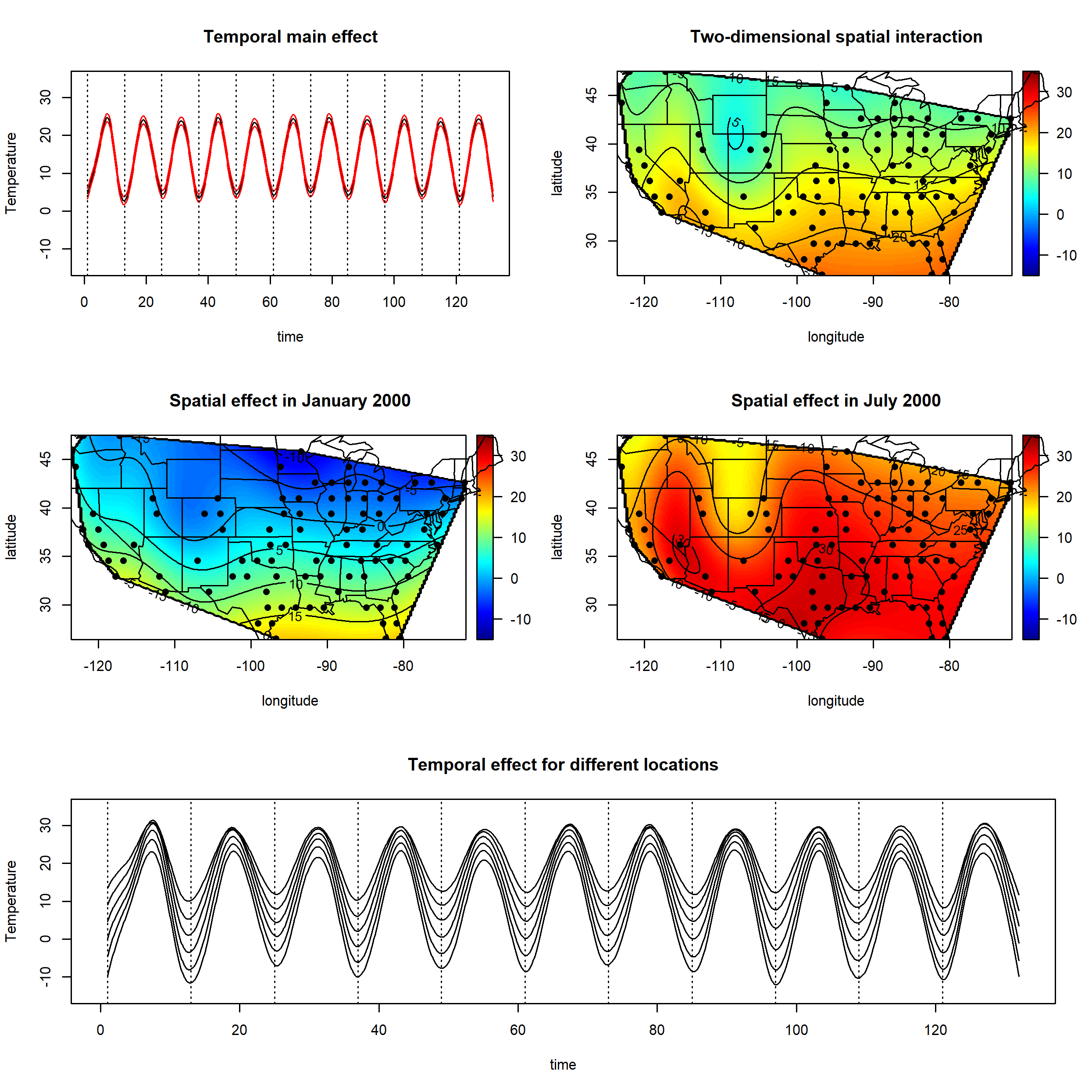}
     \caption{The plots in the first row show the estimated temporal main effect with simultaneous $95\%$ posterior credible intervals (left) and the two-dimensional spatial interaction of longitude and latitude (right). The remaining plots are slice plots. In the middle row, we fix the time (January 2000, July 2000) and plot the spatial effect. In the final plot, we fix the location (longitude=-95, latitude=30,33,36,39,42,45) and plot the temporal effect. The dashed vertical lines in the first and last plot indicate the month January for each of the years 2000 to 2010.}
    \label{fig:EffectsWB}
\end{figure}

\begin{figure}[htbp]
    \centering
    \includegraphics[width=0.9\textwidth]{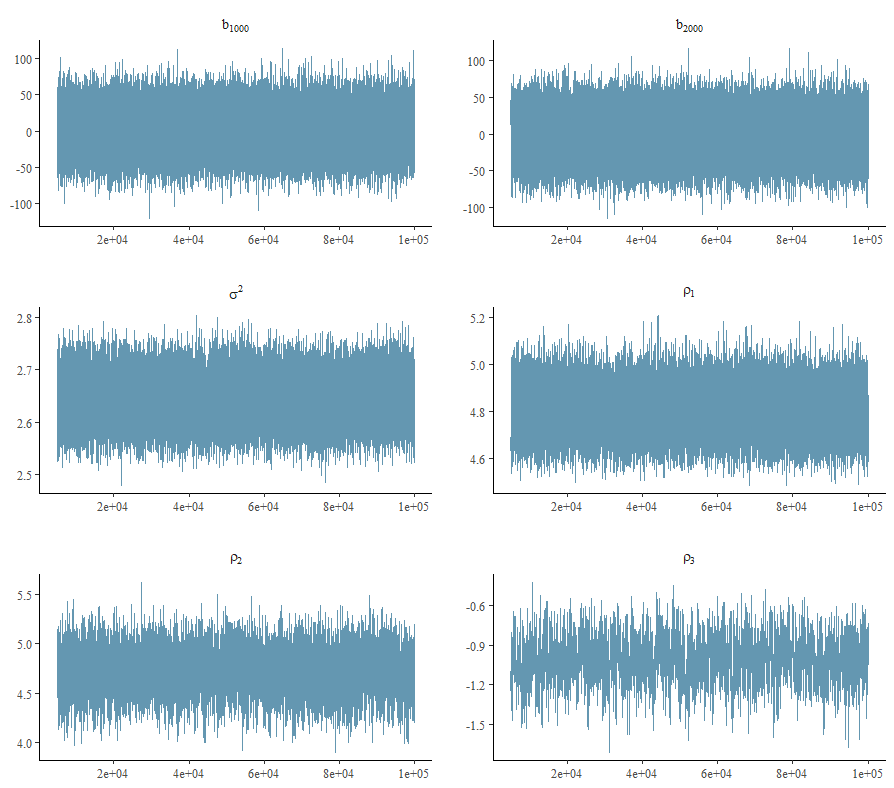}
    \caption{Shown are trace plots for two tensor product B-spline coefficients ($b_{1000}$ and $b_{2000}$), the residual variance $\sigma^2$ and the log-smoothing variances $\rho_1,\rho_2,\rho_3$.} 
    \label{fig:TracePlots}
\end{figure}

We run the MCMC sampler introduced in Section~\ref{sec:EfficientMCMC} for $T=100,000$ iterations and discard the first $5,000$ iterations as burn-in. Figure~\ref{fig:EffectsWB} shows selected functional effect estimates, while Figure~\ref{fig:TracePlots} shows trace plots for selected coefficients. Table~\ref{tab:MCMC} reports MCMC summaries as well as MCMC convergence diagnostics.

\begin{table}[htbp]
\centering
\begin{tabular}{r|lrrrrrrrrr}
 & mean & median & sd & mad & $q_{5}$ & $q_{95}$ & $\widehat{R}$ & ess (bulk) & ess (tail) \\ 
  \hline\hline
 $b_{1000}$ & 2.21 & 2.23 & 25.76 & 25.80 & -40.13 & 44.51 & 1.00 & 93588.60 & 94107.27 \\ 
  $b_{2000}$ & -1.34 & -1.28 & 25.76 & 25.77 & -43.77 & 40.93 & 1.00 & 94636.39 & 93509.84 \\ 
  $\sigma^2$ & 2.64 & 2.64 & 0.04 & 0.04 & 2.58 & 2.71 & 1.00 & 11252.77 & 37410.87 \\ 
  $\rho_1$ & 4.81 & 4.80 & 0.11 & 0.11 & 4.63 & 5.00 & 1.00 & 6667.45 & 5076.53 \\ 
  $\rho_2$ & 4.73 & 4.73 & 0.21 & 0.21 & 4.37 & 5.07 & 1.00 & 2510.79 & 4464.76 \\ 
  $\rho_3$ & -1.01 & -1.01 & 0.17 & 0.17 & -1.29 & -0.74 & 1.00 & 723.93 & 1362.43 \\

\end{tabular}
\caption{MCMC summaries and convergence diagnostics for the corresponding MCMC samples. The summaries were created using the \texttt{R} package \texttt{posterior} \citep{BurGabKay2022} and correspond to posterior means, medians, standard deviations, median absolute deviations, 5\% and 95\% quantiles, an improved version of the Gelman-Rubin $\widehat{R}$ as well as two versions of the effective sample size, one for the bulk of the distribution and one for the tails \citep[see][for further details]{VehGelSim2021}.}
    \label{tab:MCMC}
\end{table}

\subsubsection*{Conclusions}
\begin{itemize}
    \item We see that the results in the present setting are very sensible and provide interesting insights. From Figure~\ref{fig:EffectsWB} we can e.g.~see that the Rocky Mountains have a strong effect on the temperature and that the effect of the seasons is more pronounced in the north of the USA than in the south. The latter finding could not be established using an additive model of the form $f_{1}(time)+f_{2}(longitude,latitude)$. This demonstrates the advantage of the more complex model~\eqref{SpaceTimeModel} as it allows for a spatio-temporal interaction. From Figure~\ref{fig:TracePlots} and Table~\ref{tab:MCMC} we can see that the posteriors of the log-smoothing variances differ significantly, which underlines the need for anisotropic estimation.
    \item  The runtime is acceptable with 100,000 MCMC iterations taking less than nine hours (rstanarm, for comparison, has not even finished the warm-up phase of $200$ iterations by then). Moreover, the MCMC mixing for the smoothing parameters is reasonably good. \cite{VehGelSim2021} recommend that $\widehat{R}$ should be less than $1.01$ and that the effective sample size should exceed $400$ which are both satisfied (see Table~\ref{tab:MCMC}). The MH acceptance rate for the vector $\rho=(\rho_1,\rho_2,\rho_3)$ was about $64\%$.
\end{itemize}

\paragraph{Overall summary.} In summary, the temperature data example shows that our new approach is very well applicable to analyze real data. Through the visualization of lower-dimensional effects, the method allows us to gain interesting insights into complex multidimensional functions.
 
\section{Discussion}\label{sec:Discussion}
In this paper, we introduce a highly efficient fully Bayesian approach for anisotropic multidimensional smoothing using Bayesian tensor product P-splines. The key feature of our new approach are efficient and adaptive MH updates for the log-smoothing variances. These updates are possible because of the representation~\eqref{SpecialRepresentation} of the overall roughness penalty matrix, which relies on the Kronecker sum structure of the penalty matrix. We have shown that the new approach outperforms previous suggestions in the literature and demonstrated the applicability through a real data example from spatio-temporal statistics. Possible extensions are the following:

\begin{itemize}
\item Additive models: We have focused on the $p$-dimensional nonparametric regression model~\eqref{NonparametricRegressionModel} but it is straightforward to embed a $p$-dimensional tensor product smooth into a larger additive predictor. In this case, it is beneficial to introduce centering constraints for the tensor product smooth, one may e.g.~use empirical centering constraints of the form $\sum_{i=1}^n f(x_i)=0$ \citep[cf.][]{LanUmlWec2014}. The centering constraints can easily be realized in the MCMC sampler through conditioning by Kriging \citep[][Section 2.3.3]{RueHel2005}. Crucially, the update of the smoothing parameters is not affected by the centering constraints so that the approach introduced in Section~\ref{Sec:MHUpdates} can easily be carried over.
\item Non-Gaussian response models: We have focused on a Gaussian response model but our approach can easily be carried over to non-Gaussian response models. Bayesian P-splines have often been applied in additive non-Gaussian models using IWLS proposals for the B-spline coefficients \citep[see, e.g.,][]{BreLan2006,KleKneLan2015,KleKneLan2015b}. In the present setting, one can use (blockwise) IWLS proposals for the tensor product B-spline coefficients $b\in\dsR^D$.  
Crucially, the update of the log-smoothing variances $\rho$ does not depend on the likelihood so that the approach introduced in Section~\ref{Sec:MHUpdates} can directly be carried over to a non-Gaussian response setting. 
\end{itemize}